# Laboratory tests for MIR light detection and transport with specialty optical fibres

**R. Benocci**[a,b,1], **M. Bonesini**[a,c], **K.S. Gadedjisso-Tossou**[d, e, f], **H. Cabrera**[d,f], **L. Stoychev**[d,f], **M. Rossella**[g], **M. Baruzzo**[d], **M. Consonni**[c] and **J. J. Suarez-Vargas**[d,h]

[a] *Sezione INFN Milano Bicocca*
   *P.zza della Scienza 3, 20126 Milano, Italy*
[b] *Dipartimento di Scienze dell'Ambiente e della Terra, Università di Milano Bicocca*
   *P.zza della Scienza 1, 20126 Milano, Italy*
[c] *Dipartimento di Fisica "G. Occhialini", Università di Milano Bicocca*
   *P.zza della Scienza 3, 20126 Milano, Italy*
[d] *Sezione INFN Trieste*
   *Via Valerio 2, 34149, Trieste, Italy*
[e] *Département de Physique, Université de Lomé, Togo*
[f] *The Abdus Salam International Centre for Theoretical Physics*
   *Strada Costiera 11, 34151 Trieste, Italy*
[g] *Sezione INFN Pavia*
   *Via A. Bassi 6, 27100 Pavia, Italy*
[h] *Department of Mathematics and Informatics, University of Udine*
   *Via Palladio 8, Udine 33100, Italy*

   *E-mail:* `roberto.benocci@unimib.it`

ABSTRACT: The FAMU (Fisica degli Atomi Muonici) experiment at the RIKEN RAL pulsed muon facility will measure the proton Zemach radius with high precision, thus contributing to the solution of the so-called "proton radius puzzle". The situation is now confused and new measurements will help reduce the discrepancy as measured with electrons or muons. To this aim, FAMU will make use of a high-intensity pulsed muon beam at RIKEN-RAL impinging on a cryogenic hydrogen target and a tunable Mid-infrared (MIR) laser emitting at ~6.78 μm, to measure the hyperfine (HFS) splitting of the 1S state of muonic hydrogen. The injection of light into the cryogenic target and its monitoring via dedicated MIR sensors is an asset for the experiment. A possible solution is via specialty MIR fibres based on hollow core waveguides or polycrystalline fibres. Two dedicated setups based on pulsed QCL lasers, from Alpes Laser, emitting around 6.78 μm have been used to study the fibre attenuation while the high energy laser developed at Elettra Sincrotrone in Trieste has been used to study the damage threshold at high energies. The alignment of MIR laser radiation into an optical cavity presents criticalities due to geometrical constraints. Its control requires the use of small and sensitive detectors. The results obtained with a quantum IR sensor with an integrated circuit for signal processing are reported.

---

[1] Corresponding author.

**Contents**

**1. Introduction**

The recent Lamb shift experiment at PSI and the controversy about the proton size [1] raised the interest in the hyperfine splitting measurements of muonic hydrogen [2]. This measurement allows the extraction of the Zemach radius of the proton, which is the only observable parameter that characterizes both its charge and magnetic distributions. The FAMU (Fisica degli Atomi Muonici) project will be the first experiment aimed at measuring the hyperfine splitting (HFS) in the muonic-hydrogen atom ground state. Its experimental method requires a detection system suited for time resolved X-ray spectroscopy using HPGe detectors and scintillating counters based on LaBr3(Ce) crystals [3]. The FAMU experiment will be performed at the RIKEN RAL facility where a high intensity pulsed muon beam is available. The interaction of the muon beam with a hydrogen gas mixture produces muonic hydrogen (μp) atoms. In subsequent collisions with $H_2$ molecules, the μp de-excites at thermalized μp in the $(1S)_{F=0}$ state. Afterwards, the muon from muonic hydrogen is transferred to the nucleus of a heavier gas added to the hydrogen (such as oxygen) and such transfer is detected by observing the characteristic relaxation X-rays spectrum. This process occurs at typical time scale of 300 ns after the μp formation. If in the time interval between the μp thermalization and the muon transfer to oxygen, the muonic hydrogen atoms absorb a photon at the resonance frequency provided by a laser source, a hyperfine para-to-ortho transition $((1S)_{F=0} \rightarrow (1S)_{F=1})$ is promoted and, in the following collisional de-excitation to the para spin state, they are accelerated by ∼ 0.12 eV [4]. The resonance frequency photon is provided by a tunable high energy MIR laser developed at Elettra Sincrotrone Labs [5]. The number of atoms that have undergone the above sequence can be determined by observing a change of the time spectrum distribution of the characteristic X-rays emitted after the transfer of the muon from the muonic hydrogen to the nucleus of oxygen added to the hydrogen target. Indeed, the transfer rate is an energy-dependent process in the epithermal energy range. The specific wavelength used is 6785±10 nm. One of the main drawbacks of using such wavelength is its high absorption in air due to the presence of water vapour. Therefore, the major concern related to the laser light transport from the laser source to the gas target cavity are the laser losses. One possibility is the use of a vacuum tube or , as an alternative, a vapour-free gas pipeline. In this case, one of the possible disadvantages may arise from alignment difficulties due to geometrical constraints and the possible surge of vibrations as the two systems, laser and target, are decoupled. Therefore, a possible solution may come from the use of fibre optics transparent to MIR radiation. This arrangement would have the advantage of avoiding the building up of a dedicated beam transport channel with the associated alignment issues. However, in the initial stages of the experiment the "direct injection" has been preferred. In this case, the alignment of the MIR laser beam will be controlled by the use of small and sensitive detectors. The results obtained with a quantum IR sensor with an integrated circuit for signal processing are reported.

Mid-Infrared (MIR) radiation around 6-7 μm has also important applications for biomedical diagnostics, environmental monitoring, isotope analysis. It is known, for example, that any disease process is preceded by changes in the metabolism of the affected cells or organs. Thus, new technologies such as Fourier transform infrared spectroscopy (FTIR) has emerged as a valuable alternative to detect bio-molecular changes in the fields of biology and medicine yielding information on a biologic state in early medical diagnosis [6, 7]. Infrared optical fibres based on chalcogenide glass have been used as sensor for the determination of volatile organic pollutants in groundwater motivated by the lack of fast and cost-efficient state-of-the-art techniques on the one hand and the increasing demands of national and international directives concerning the quality of water on the other [8, 9]. Fibre-optic components show enhanced versatility with respect to bulk optics, mostly in terms of flexible beam guidance and compactness. Furthermore, fibres could also be deployed advantageously as modal wave-front filters, optical path delay length control or for multi-axial beam combining [10]. There are various fields of application for infrared fibres, each one requiring specific fibre properties such as: nulling interferometry [11], high precision imaging and spectroscopy [12].

## 2. Tests of optical fibres in MID-IR range

The performance of an optical fibre for the transport of a specific radiation is determined evaluating the attenuation, A, or the attenuation coefficient, α, that is the attenuation per unit length expressed in dB:

$$A = 10 \log \left(\frac{P_{ref}}{P_{out}}\right) \qquad (1)$$

$$\alpha = \frac{10}{z} \log \left(\frac{P_{ref}}{P_{out}}\right) \qquad (2)$$

where $P_{ref}$ is a reference power such as the input power into the fibre, $P_{out}$ is the output power from the fibre and z is the fibre length. In the literature [13], the optical fibres used for the transport of radiation in the range 6-7 um are: polycrystalline Silver Halide fibres, Hollow glass fibres and Chalcogenide glass fibres. The reported attenuation coefficient, α, is of the order of 1 dB/m. Therefore, the following optical fibres have been selected for the initial tests:

- Hollow Fibre HF500MWLW-SMA-1m ID = 500 μm; MWLW Silver Iodide coating for λ = 5 – 12 μm; SMA connectors, from Opto-Knowledge Systems
- RF-Se-300 multimode LW mid-infrared Chalcogenide glass fibre for λ =1.5 - 9.7 μm, ID =300 μm, FC connectors, from IRFLEX
- PIR AgCl:AgBr Polycrystalline fibre, cores of 410 μm and 860 μm, from ArtPhotonics

Two are the main requirements the selected optical fibres have to meet in order to be implemented in the laser transport line:
1. Present a low attenuation coefficient in the wavelength range 6.78-6.80 μm,
2. Have a damage threshold capable to withstand the high energy laser at around 6.80 μm (energy of about 1 mJ with pulse duration of 10 ns).

For this purpose, different measurement campaigns have been performed to address this issue. In particular, two QCL lasers have been used to investigate the fibres attenuation with different operative and injection conditions, whereas a Difference Frequency Generation (DFG) based high energy laser beam has been used to study the damage threshold of the different fibre materials.

## 2.1 Tests with cryogenic pulsed QCL laser

The layout of the measurement system is shown in Fig. 1. We used a pulsed Distributed-feedback Quantum cascade laser (DFB-QCL, #sb 1486 DN, Alpes Lasers Technology). The QCL was operated with pulse duration of 100 ns, repetition rate of 200 kHz. The operation temperature was changing from 80 K to 150 K allowing to have a single mode operation in the wavelength range between 6780 and 6825 nm with a line width less than 1 nm. Operating the laser at this duty cycle (2%) allows to reach an average power of 12 mW. The beam from the laser was collimated by a Ge lens of 25.4 mm focal length ($L_1$) and then focused by a ZnSe lens of 40 mm focal length ($L_2$) on a SMA/FC adapter after two reflexions on gold coating mirrors $M_1$ and $M_2$. The reference fibre (orange jacket) was connected to the fibre to be tested (green jacket) by a mating adapter (SMA) and the output light was collected by a power meter via a SMA adapter (Fig. 1).

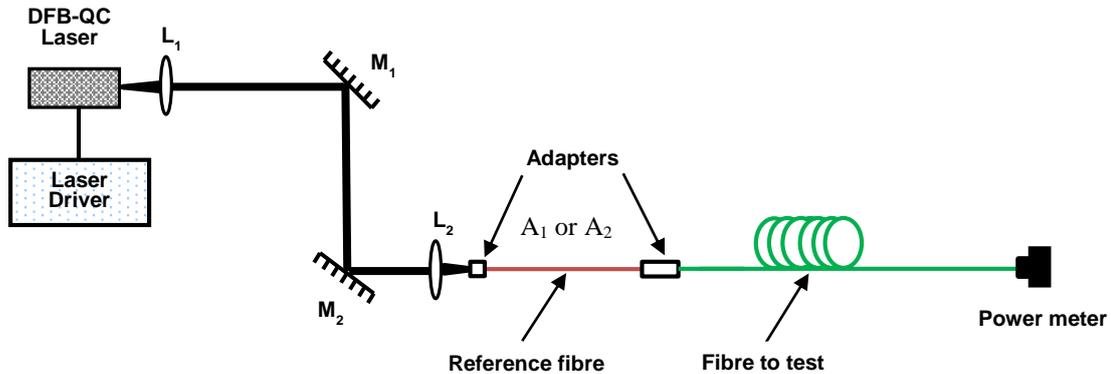

**Figure 1.** Experimental arrangement for the attenuation measurement with a pulsed DFB- QCL laser.

For the measurements we proceeded as follow:
1. We measured the output power from the SMA ($A_1$) or FC ($A_2$) fibre adapter,
2. The reference fibre was connected between $A_1$ or $A_2$ adapter and the power meter, and a first measurement is taken, $P_{ref}$.
3. Afterword, the fibre under test was connected between the reference fibre and the power meter and a second measurement is taken, $P_{out}$.
4. We connected another fibre to the previously mounted fibre and we measured the output power.

At our disposal we had two 50 cm long fibres and one 1 m long fibre for both the hollow and chalcogenide glass fibre. For the Silver Halide we had just a 2 m long fibre. With this procedure both the coupling and the transmission/attenuation factor could be evaluated. We denoted such value as total losses. The insertion loss due to the mating fibre adapter has been measured by comparing for example the output power from a 1 m long fibre and two 50 cm long fibres. As detector we used an Ophir power meter (model 3A-SH; thermopile; spectral range: 0.19-20 μm; power range: 10 μW-3 W) and a Kolmar detector (model KMPV11-0.5-J2; HgCdTe photodiode. peak cut-off wavelength: 10.5 μm; sensing area: 0.5x0.5 mm).

**2.2 Tests with pulsed pigtailed QCL laser**

The laser used in this second set of measurements is a Pulsed Pigtailed HHL-696 Alpes Laser QCL with a Thermo-Electric Cooler (TEC) system in order to reach -40°C. The temperature control allowed to scan the laser emission between 6780 to 6816 nm by changing the operative temperature from +20 C to -40 C. The emission power is almost constant over the range of wavelength with an average power at the maximum supplied current (1.35 A) of around 3.5 mW. The employed laser has the following features:
- Pulse duration of 20 ns and repetition rate of 1MHz.
- Single mode with line width <1.3 nm.

A CaF lens with 20 mm focal length is directly mounted on the laser base together with a SMA fibre holder adapter as it can be seen in Fig. 2.

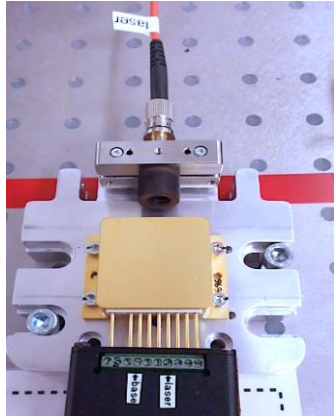

**Figure 2.** Pigtailed QCL Laser head mounted on a water cooled base, pulser and fibre plus lens holder.

In this case, we performed two kind of measurements to evaluate both the attenuation coefficient and the total losses due to attenuation and injection. For the measurement of the attenuation coefficient, we proceeded as follow:
1. A 50 cm long fibre is connected directly to the power meter and a first power measurement is taken ($P_{ref}$).
2. Afterword, we changed the fibre with a 1 m long fibre and a second measurement is taken ($P_{out}$).

The attenuation coefficient in then calculated by Eq. (1). For the evaluation of the total losses, we proceeded as follow:
1. The emitted power from the laser is taken after the lens/SMA adapter ($P_{ref}$)
2. Afterword, we connect a 50 cm long fibre and a second measurement is taken ($P_{out}$).

The total losses are then calculated by Eq. (2).

**2.3 Tests for the evaluation of the damage threshold**

The laser, developed in collaboration with the Elettra Sincrotrone Labs in Trieste, has to satisfy strict requirements to be of interest for the FAMU experiment in order to excite an efficient spin flip transition of the muonic hydrogen. The laser system is divided into two main units.
1. The Cr:Fo master-oscillator power- amplifier (MOPA) [8]. It is made of a Cr:Fo based oscillator delivering a very narrow linewidth (0.5 pm), single longitudinal mode (SLM) pulsed laser light centred at 1262 nm and a four-stages amplifier unit. The total output energy is 24 mJ at 25 Hz.

2. The second subsystem is a Nd:YAG, laser with 18 ns pulse duration, jitter 0.5 ns, 25 Hz repetition rate, and a maximum energy output < 300 mJ. This laser is seeded with a highly stable fibre laser at 1064 nm.

These two laser wavelengths are combined in a LiInS2 crystal (NLC) with dimensions 7 x 7 x 20 mm$^3$, to perform DFG. The DFG process gives access to a broad MIR region 6445 – 6935 nm with an efficiency of 1.4% with respect to the energy of the Nd:YAG pump (560 µJ with 39 mJ from the Nd:YAG). In our case, we have been able to produce up to 140 µJ in single pass and 560 µJ in double pass configuration. The scheme of the laser system for DFG is shown in Fig. 3.

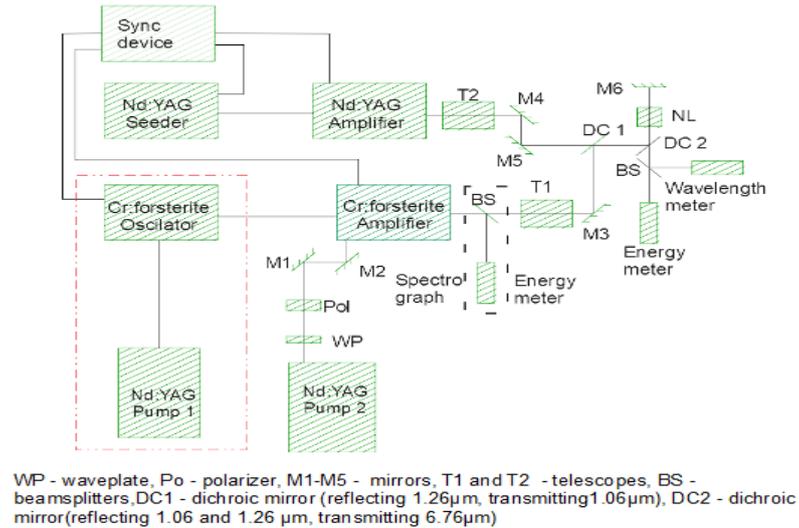

**Figure 3.** Laser scheme at Elettra Sincrotrone Labs in Trieste.

## 3. Results ad Discussion

In this section, we present the results obtained using the three lasers described in Sect. 2.1, 2.2, 2.3.

### 3.1 Transmission tests

In Fig. 4, the total losses (i.e. the combination of both the attenuation and laser to fibre coupling) as a function of the fibre length for three different fibres are illustrated. For this measurement we used the cryogenic pulsed QCL laser with the emission wavelength set at λ=6789.3 nm. As expected, the attenuation increases by increasing the length of fibres, with the attenuation for the hollow fibre always higher than the chalcogenide glass fibre. During this set of measurements, we had at our disposal a 2 m long Silver Halide fibre with an attenuation slightly above 5 dB.

Figure 5 shows the total losses as a function of the focusing length for the three fibres (50 cm long: hollow and chalcogenide glass fibres; 200 cm long: Silver Halide fibre). As it can be clearly seen, the hollow fibre presents a minimum of attenuation when using a lens with 150 mm focal length. The other fibres do not seem to be affected by such injection parameter. Also in this case, the Silver Halide fibre presents lower values of attenuation. A possible explanation for this behaviour can be the following: although coupling light into a hollow fibre is relatively simple given the large core, both transmission and beam quality can be adversely affected if the proper focal length optic is not used. In general, the beam should enter with a small angle into the fibre that is with a relatively gradual focus. Optimal coupling into the lowest order mode occurs when the ratio of the focused spot size to the fibre internal diameter, ID, is $2\omega_0/ID = 0.64$ [15, 16], where $\omega_0$ is the beam waist. In our case, for ID = 500 μm and a laser beam diameter 5 mm we get a focal length of about 160 mm. All measurements in Figs. 4-5 have been corrected by removing the attenuation due to the insertion loss caused by the use of fibre mating adapter. Its contribution has been estimated for different laser emission wavelengths. Mean insertion loss values of 0.8, 1.5 and 2 dB have been obtained for the chalcogenide glass, Silver Halide fibres with 400 μm and 800 μm, respectively.

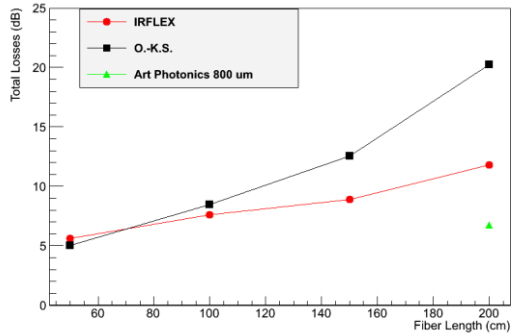 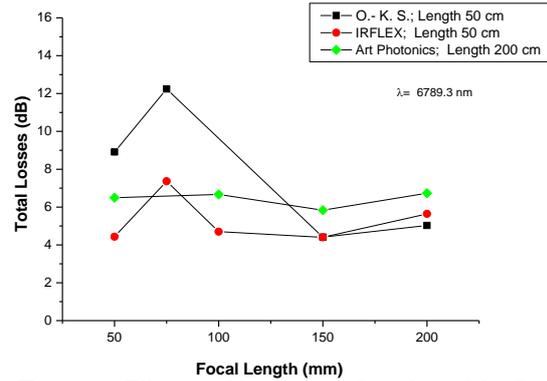

**Figure 4**. Total losses as a function of the fibre length for three different fibres: Hollow fibre (O.-K.L.), Chalcogenide glass fibre (IRFLEX) and Silver Halide glass fibre (Art Photonics). Focusing lens: Ge=200 mm. The laser emission wavelength is λ=6789.3 nm. Tests performed using the cryogenic pulsed QCL laser.

**Figure 5**. Fibre total losses as a function of the focal length for the three tested fibres. In this case, the fibre length is 50 cm for the Hollow fibre (O.-K.L.) and Chalcogenide glass fibre (IRFLEX), and 200 cm for the Silver Halide (Art Photonics). Tests performed using the cryogenic pulsed QCL laser.

In Fig. 6, the total losses are displayed as a function of the laser emitting wavelength. These are obtained by comparing the output power measured at the laser exit and the power measured after 50 cm long fibre. This and the following measurements have been performed using the pulsed pigtailed QCL laser. As it can be clearly seen, the lowest losses are achieved for the Silver Halide glass fibre (Artphotonics) with 800 μm core with values of about 1.5 dB in the whole range of wavelengths. Larger core diameter thus provides less attenuation, most likely due to a better laser-fibre coupling. Conversely, the performance of the hollow fibre presents very high losses (>6 dB) with a high uncertainty especially in the wavelength range 6787-6796 nm.

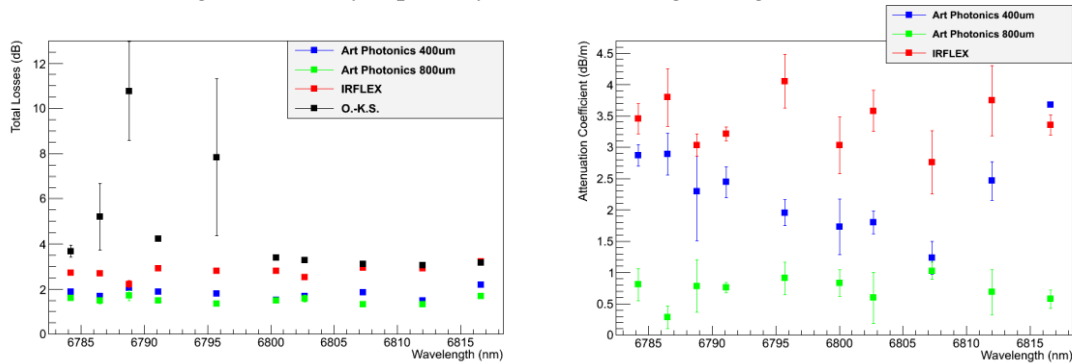

**Figure 6.** (Left panel) Total losses due to the laser-fibre coupling obtained by comparing the output power measured at the laser exit and the power measured after 50 cm long fibre. (Right panel) Attenuation coefficient obtained by comparing the output power from a 50 cm long fibre and the 1 m long fibre. Tests

performed using the cryogenic pulsed QCL laser.

As a comparison, we report in Fig. 7 the output power at the exit of a 50 cm long hollow and Silver Halide glass (ID 800 μm) fibre as a function of the laser wavelengths and at different supplied powers. Since the power transmitted by the hollow fibre around 6789 nm and at 6796 nm is nearly negligible, the total losses corresponding to such values are very high as already reported in Fig. 6 (Left panel). The sudden drop around these two wavelength (also partially present for the Silver Halide fibre: right panel of Fig. 7) is due to the absorption by water vapour in air. Figure 8 reports the result of the transmittance simulation of MIR light between 6710 and 6860 nm at a pressure of 1 bar, a temperature of 296 K and an optical path of 1 cm using the *high*-resolution *tran*smission molecular absorption database (HITRAN) [17].

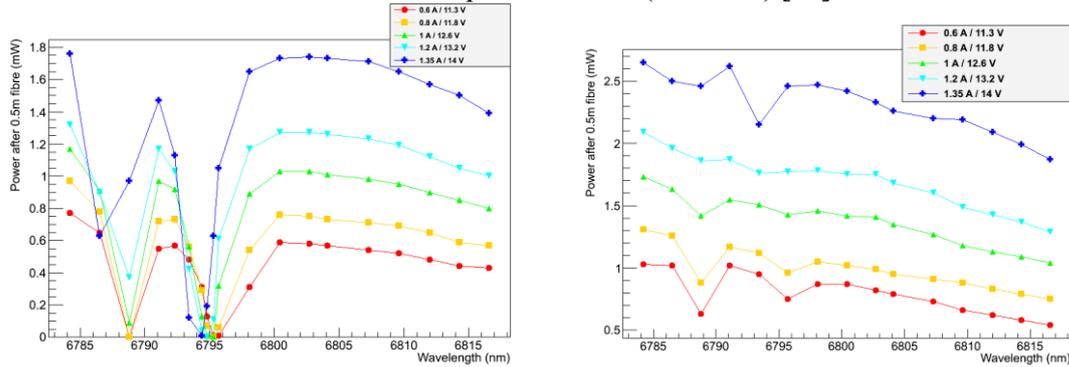

**Figure 7.** Power measured at the exit of a 50 cm long hollow fibre (Left panel) and at the exit of a 50 cm long Silver Halide glass fibre with ID 800 μm (Right panel) as a function of the laser wavelength and at different supplied powers. Tests performed using the cryogenic pulsed QCL laser.

Figure 8 gives two transmittance minima at about 6787 and 6793 nm which correspond fairly well with those found experimentally at 6789 and 6794/5 nm. Thus, hollow fibres provide an additional absorption at those wavelengths due to both the presence of water vapour in the core and adsorbed on the core surface.

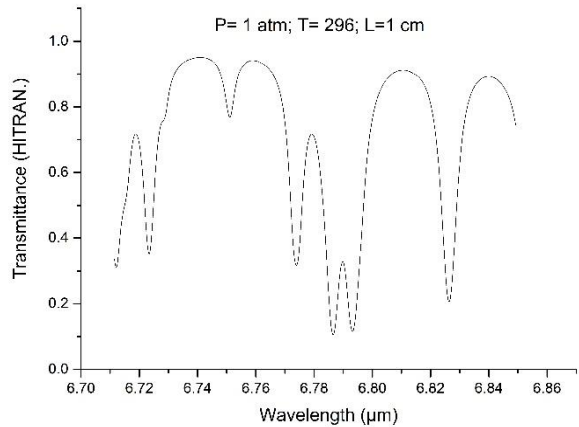

**Figure 8.** Transmittance of MIR light between 6710 and 6860 nm at a pressure of 1 bar, a temperature of 296 K and an optical path of 1 cm using the high-resolution transmission molecular absorption database (HITRAN).

**3.2 Damage tests**

It is known that for an ideal surface, the optical damage threshold at the surface is close to that of the bulk medium. In practice, the damage threshold at the surface can be much lower than that of the bulk because of the presence of defects and contaminations. For example, it has been shown that the optical damage threshold at 1064 nm is close to 480 GW/cm$^2$ in bulk silica due to electron avalanche and independent of pulse width for pulse durations above 100 ps [18]. For shorter pulses, the avalanche evolves more slowly than the pulse envelope, leading to a slightly higher threshold [18]. In our case, the damage threshold of MIR fibres has never been tested at the specific wavelength around 6785 nm using a 10 ns high energy pulsed laser. The peak energies, the optical fibres have been exposed to, in the tests described in Sect. 2.1 and 2.2 were of the order of tens of nJ. At the Elettra Sincrotrone Labs in Trieste, the laser has been designed to deliver energies which are in the range of mJ. Therefore, our purpose was to test the capability of these MIR fibres to transport the laser light at such high energies observing any eventual damage threshold. At the present stage, the laser can deliver up to about 200 µJ.

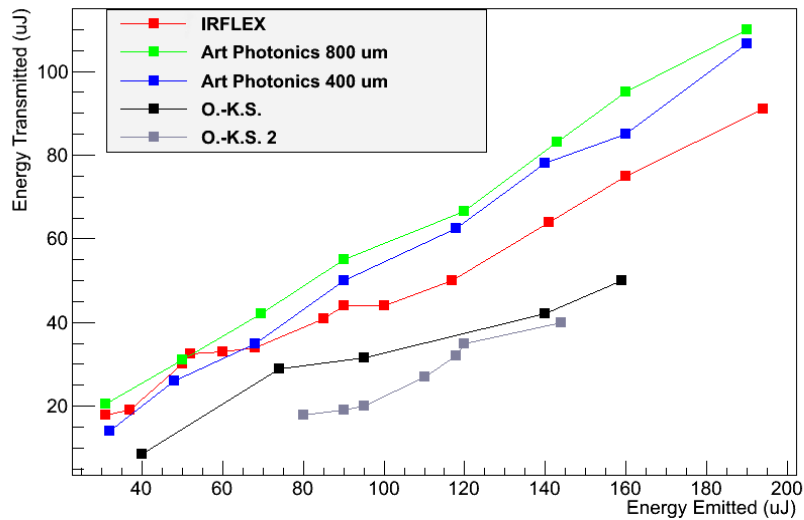

**Figure 9.** Energy transmitted by the different optical fibres; for the OptoKnowledge fibres (hollow) the subscripts 1 and 2 refers to different focusing lengths: "1" CaF 150 mm, "2" CaF 200 mm. For the other fibres a Ge 200 mm focusing lens has been used. Laser wavelength λ= 6787.51 nm.

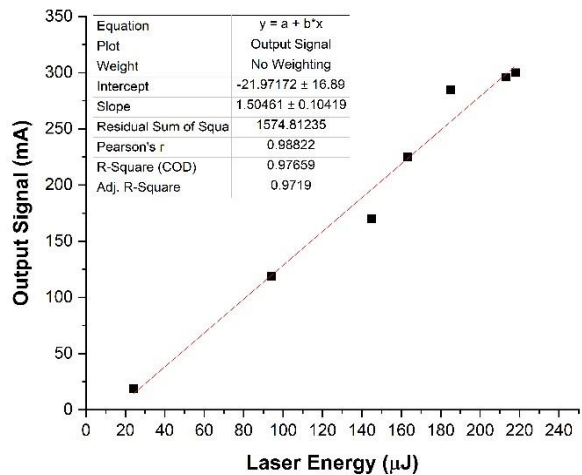

**Figure 10.** Sensor's output signal as a function of laser energy. The range of energy spans between 24 and 218 µJ. The red line represents a linear fit (R-Square =0.977).

Figure 9 shows the results of the energy transmitted by the different fibres as a function of laser light energy, obtained at a laser wavelength λ= 6787.51 nm. The energy transmitted is linear in the range 20-200 µJ showing no damage to the fibres at peak power densities of about $10^7$ W/cm$^2$. The total losses obtained in these tests confirmed those obtained in previous tests (Sect. 3.1) including the tests on the hollow fibres that shows a higher transmittance when the light is focused using 150 mm focal length. During the damage tests, we also investigated the behaviour of a MIR sensor developed for detecting human presence. This is justified by the need to align the laser beam in an experimental setup with strong geometrical constraint [19, 20]. We used the AK9752 module, a very low power and small (2.2 mm x 2.2 mm) IR sensor. It is composed of a quantum IR sensor and an integrated circuit for signal processing. An integral analog-to-digital converter provides 16-bits data outputs [21]. These characteristics make this detector suitable to be employed in the alignment operation of the laser beam in the FAMU experiment. We tested the response to the laser light at different energies using the beam line of the Elettra laser. The response of the sensor between 24 and 218 µJ is linear with a R-Square (statistical measure that represents the proportion of the variance for a dependent variable that is explained by an independent variable or variables in a regression model [22]) of about 0.98, as it can be observed in Fig. 10.

## 4. Conclusions

The use of optical fibres to convey the MIR laser beam at 6785 nm has been addressed in this paper. Three different fibres have been investigated: a hollow fibre and two polycrystalline glass fibres. The results showed that Silver Halide glass fibres perform better than the other two fibres and that increasing the core diameter the attenuation is reduced. In particular, we found that the Silver Halide glass fibre with a core diameter of 800 µm provided an attenuation coefficient <1 dB/m and total losses (accounting for the attenuation and injection losses) for a 50 cm long fibre of about 1.5 dB. No damage threshold is present up to about 200 µJ. Further tests are necessary to extent the range of the threshold to energies > 1-2 mJ. These results make the Silver Halide glass fibres promising for their implementation in the FAMU experiment, thus significantly reducing the criticalities associated with the "direct injection".

The alignment of MIR laser radiation in the FAMU experiment is not an easy task due to the specific laser light wavelength and the geometrical constraints which demand for small and sensitive detectors. The results obtained for a quantum IR sensor with an integrated circuit for signal processing revealed that the sensor response is linear within 24 and 218 µJ with a R-Square of 0.97 and therefore suitable for the FAMU alignment process.